\documentclass[amsmath,amssymb,aps,prl,nofootinbib,superscriptaddress,twocolumn,preprintnumbers]{revtex4-2}
\usepackage[utf8]{inputenc}
\usepackage{mathrsfs}
\usepackage{bm}
\usepackage{url}
\usepackage[normalem]{ulem}
\usepackage{mathtools}
\usepackage{array}
\newcolumntype{P}[1]{>{\centering\arraybackslash}p{#1}}
\newcolumntype{M}[1]{>{\centering\arraybackslash}m{#1}}
\usepackage[caption=false]{subfig}
\usepackage{dcolumn}
\usepackage{graphicx, epsfig}
\usepackage[dvipsnames]{xcolor}
\usepackage{mathrsfs}
\usepackage{bm}
\usepackage{yhmath}
\usepackage[caption=false]{subfig}
\usepackage[normalem]{ulem}
\usepackage{mathtools}
\usepackage{bigints}
\usepackage{float}
\usepackage[colorlinks = true, linkcolor = purple, urlcolor  = blue, citecolor = blue, anchorcolor = blue]{hyperref}
\usepackage{float}
\usepackage{multirow}
\usepackage{tikz,xcolor,hyperref}
\definecolor{darkgreen}{rgb}{0.0, 0.2, 0.13}
\definecolor{bostonuniversityred}{rgb}{0.8, 0.0, 0.0}
\definecolor{lime}{HTML}{A6CE39}
\DeclareRobustCommand{\orcidicon}{
	\begin{tikzpicture}
	\draw[lime, fill=lime] (0,0) 
	circle [radius=0.16] 
	node[white] {{\fontfamily{qag}\selectfont \tiny ID}};
	\draw[white, fill=white] (-0.0625,0.095) 
	circle [radius=0.007];
	\end{tikzpicture}
	\hspace{-2mm}
}
\foreach \x in {A, ..., Z}{\expandafter\xdef\csname orcid\x\endcsname{\noexpand\href{https://orcid.org/\csname orcidauthor\x\endcsname}
			{\noexpand\orcidicon}}
}
\newcommand{\be}{\begin{equation}}
\newcommand{\ee}{\end{equation}}
\newcommand{\ba}{\begin{eqnarray}}
\newcommand{\ea}{\end{eqnarray}}

\def\lp4{$\lambda \phi^4$}
\begin{document}
\preprint{\texttt{FERMILAB-PUB-26-0229-T}}
\title{High-energy neutrino constraints on primordial black holes as dark matter}
\author{Mainak Mukhopadhyay\hspace{-1mm}\orcidA{}}
\email{mainak@fnal.gov}
\affiliation{Astrophysics Theory Department, Theory Division, Fermi National Accelerator Laboratory, Batavia, Illinois 60510, USA}
\affiliation{Kavli Institute for Cosmological Physics, University of Chicago, Chicago, Illinois 60637, USA}
\author{Joaquim Iguaz Juan\hspace{-1mm}\orcidA{}} \email{jiguazjuan@umass.edu}\thanks{Equal contribution.}
\affiliation{Department of Physics, University of Massachusetts, Amherst, MA 01003, USA
}
\date{\today}
\begin{abstract}
%
%
Primordial black holes (PBHs) are one of the most appealing dark matter candidates over a wide range of masses and abundances. This broad parameter space has been constrained by a variety of observational probes. In this work, for the first time, we use data from high-energy neutrino telescopes, like IceCube and ANTARES, to constrain sub-asteroid mass ($\lesssim 10^{18}\,\mathrm{g}$) Schwarzschild PBHs with extended mass functions. We derive limits from the diffuse high-energy neutrino flux produced by the direct evaporation of PBHs, as well as from the transient signatures associated with PBHs passing in the vicinity of the Earth. While our bounds are slightly weaker than existing constraints from gamma-ray observations, they provide an independent and complementary probe based on observational high-energy neutrino data. We further show that future detectors such as IceCube-Gen2 and KM3NeT can significantly improve these constraints, potentially excluding PBHs with masses up to $\sim \mathrm{few} \times 10^{18}\,\mathrm{g}$ composing the entirety of dark matter.
\end{abstract}
\maketitle
\begin{figure}
\centering
\includegraphics[width=0.95\textwidth]{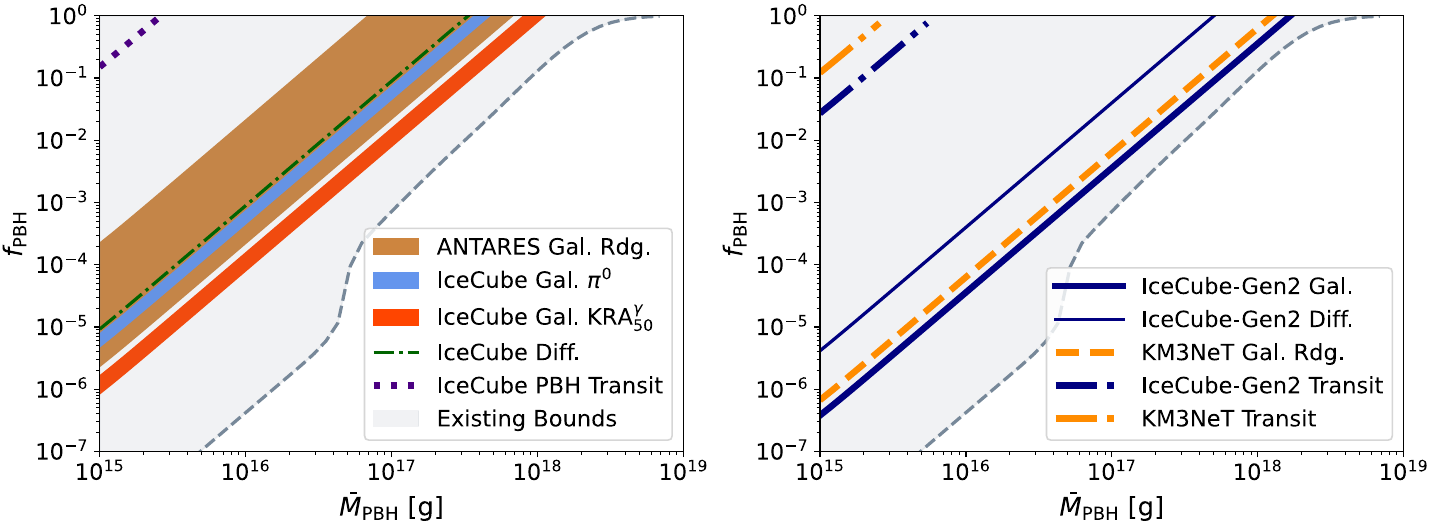}
\caption{\label{fig:res}Constraints derived in this work using high-energy neutrino observations in current detectors \emph{(top)} and predictions for upcoming neutrino detectors \emph{(bottom)} for Schwarzschild PBHs with an extended GCC mass function. The existing most constraining bounds~\cite{Berteaud:2022tws,Korwar:2023kpy} are shown after scaling for the GCC mass function for $\alpha = 1$ and $\beta = 2.78$~\cite{Carr:2017jsz}. The thick bands for the ANTARES and IceCube limits for the Galactic ridge and plane respectively denote the $90\%$ C.L. The IceCube diffuse limit is obtained from the best-fit broken power-law model. The limits are derived by requiring that the PBH-induced flux does not exceed the observational bounds at any energy (see Eq.~\ref{eq:fpbhlim}), and are therefore \emph{conservative}. All PBH transit limits are obtained at a conservative $5\sigma$ C.L. in the absence of backgrounds (Northern Sky for IceCube) assuming an operation time of $12.5$ years.
}
\end{figure}
\textit{\textbf{Introduction.}}
%
The nature of dark matter (DM) remains one of the most compelling open questions in modern physics. Primordial black holes (PBHs) have long been considered a viable candidate that could contribute to the present-day DM abundance~\cite{Green:2020jor,Carr:2021bzv,Carr:2026hot}. PBHs are typically expected to form in the early universe from the collapse of large density fluctuations upon horizon reentry~\cite{Zeldovich:1967lct,10.1093/mnras/152.1.75,1974MNRAS.168..399C}. Moreover, realistic models often predict extended mass functions spanning several orders of magnitude~\cite{Choptuik:1992jv,Niemeyer:1999ak,Green:1999xm}. However, a wide variety of alternative formation mechanisms have been proposed in the literature (see, e.g.,~\cite{Carr:2017jsz,Carr:2020gox} and references therein). This diversity underscores that PBH production is not tied to a single scenario, but can arise from a broad class of early-universe processes. A distinctive feature of PBHs is that, due to their inverse mass–temperature relation, $T_{\rm BH} \propto 1/M_{\rm BH}$, they undergo Hawking evaporation and emit all particle species with masses $\lesssim T_{\rm BH}$~\cite{1974Natur.248...30H,Hawking:1975vcx}. This makes their evaporation products a promising avenue for detection. However, identifying the relevant mass ranges and determining what fraction of PBHs in those ranges can constitute the DM abundance, typically parameterized by $f_{\rm PBH}$, remain central challenges.

In recent years, considerable effort has been focused on constraining the lower end of the so-called asteroid mass gap ($10^{17}\mathrm{g} \lesssim M_{\rm PBH} \lesssim 10^{22}\mathrm{g}$), where PBHs may still constitute a significant fraction (or even the entirety) of DM. These constraints are primarily derived from searches for Hawking evaporation products, most notably gamma rays~\cite{Coogan:2020tuf,Iguaz:2021irx,Chen:2021ngo,Berteaud:2022tws,Korwar:2023kpy,DelaTorreLuque:2024qms}, with complementary probes from antimatter~\cite{Laha:2019ssq,DeRocco:2019fjq,Huang:2024xap,DeRomeri:2025dwm} and MeV neutrinos~\cite{Dasgupta:2019cae,Bernal:2022swt}. At higher masses, PBHs are too cold to evaporate efficiently, and constraints instead arise from other probes~\cite{Green:2020jor,Carr:2021bzv,Carr:2026hot}.
An important and relatively overlooked avenue is the use of neutrinos from PBH evaporation as an independent messenger, both for constraining their abundance and for identifying nearby fly-by of PBHs. While low-energy neutrinos have been investigated~\cite{Lunardini:2019zob,Dasgupta:2019cae,Bernal:2022swt}, the high-energy regime remains largely unexplored, the only study we are aware of that sets constraints on $f_{\rm PBH}$ is~\cite{Chianese:2024rsn}, in the memory-burden scenario, and~\cite{Dave:2019epr,Capanema:2021hnm,Airoldi:2025bgr} which study the detectable signal from a transit PBH event in the local neighborhood. In light of the current generation of powerful neutrino telescopes, as well as upcoming ones with improved sensitivity, it is timely to assess the constraints that high-energy neutrino observations can place on PBHs and to forecast their future reach.

Besides the above, there is yet another motivation of why it is important to look at the high-energy neutrino landscape in the context of PBHs. In February 2025, the KM3NeT collaboration reported the observation of an extraordinarily high-energy neutrino event, with an estimated energy of approximately $220\,\mathrm{PeV}$~\cite{KM3NeT:2025npi}. This marks the first detection of a neutrino at such an extreme energy, despite more than a decade of observation by IceCube. The apparent tension between the two experiments, including the short observation time of KM3NeT relative to IceCube, disfavors a transient point source explanation at the $2\sigma$ C.L.~\cite{Li:2025tqf}, which hints towards the possibility that the KM3NeT event originated from an unknown astrophysical source. Several scenarios have been proposed in the literature to explain the KM3NeT event. Among the simplest is the PBH hypothesis~\cite{Boccia:2025hpm,Klipfel:2025jql,Baker:2025cff}, which can naturally emit extremely high-energy particles if their masses are sufficiently small\footnote{The KM3NeT explanation assuming the usual Schwarzschild PBH and the Standard Model has been challenged by the non-observation of accompanying gamma rays~\cite{Airoldi:2025opo}. However, PBH scenarios beyond these assumptions still provide a satisfactory explanation compatible with gamma-ray observations~\cite{Baker:2025zxm,Baker:2025cff}.}. Given this possibility, it is crucial to understand what directions can be pursued in light of the increasingly advanced high-energy neutrino telescopes.

In this \emph{Letter}, for the first time we present high-energy neutrino constraints on PBHs derived from observational data from IceCube~\cite{IceCube:2023ame,IceCube:2024fxo} and ANTARES~\cite{ANTARES:2022izu} and investigate the landscape in the context of the upcoming high-energy neutrino telescopes like IceCube-Gen2~\cite{IceCube-Gen2:2020qha,Gen2_TDR} and KM3NeT~\cite{KM3NeT:2024paj}. We will show that even the \emph{conservative} constraints from the current data already allow us to exclude a portion of the asteroid mass gap as entirety of DM for extended PBH mass functions, while data from the next-generation telescopes can exclude up to $\sim$ a few $10^{18}$g PBHs, comparable to the leading constraints. We also investigate the potential of high-energy neutrino telescopes to observe a fly-by of a PBH to set stringent $5\sigma$ C.L. limits on the horizon distance and the resulting constraints. The novelty lies in the fact that these provide us with an \emph{independent} messenger (high-energy neutrinos) to probe at least a portion of the asteroid mass gap with plethora of observational high-energy neutrino data.

\textit{\textbf{Constraints from diffuse neutrino fluxes.}}
\begin{figure}
\centering
\includegraphics[width=0.48\textwidth]{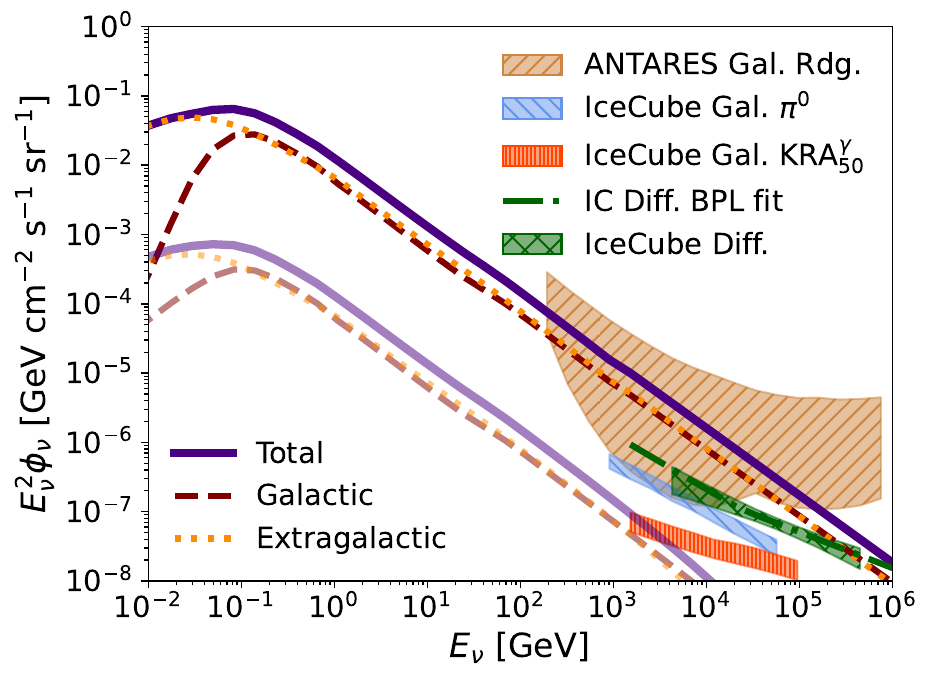}
\caption{\label{fig:gal_egal_flux}Galactic (Eq.~\ref{eq:gal_avg}), extragalactic (Eq.~\ref{eq:exgal}), and the total diffuse all-flavor neutrino contribution from Schwarzschild PBHs, including both primary and secondary neutrinos, averaged over all-sky shown separately for $\bar{M}_{\rm PBH} = 10^{15}$ g, $f_{\rm PBH} = 10^{-4}$ (darker) and $10^{17}$ g, $f_{\rm PBH} = 10^{-2}$ (lighter) for $\alpha = 1$ and $\beta = 2.78$. We also show the diffuse all-flavor neutrino contribution from the Galactic Ridge from ANTARES~\cite{ANTARES:2022izu} and the Galactic Plane~\cite{IceCube:2023ame} and all-sky from IceCube~\cite{IceCube:2024fxo} along with the broken power-law (BPL) fit.
}
\end{figure}
The diffuse neutrino emission from a population of PBHs can be split into the galactic and extra-galactic contributions. We leverage the fact that the high-energy neutrino telescopes like ANTARES has recently measured the diffuse neutrino emission from the Galactic Ridge~\cite{ANTARES:2022izu} and IceCube has searched for the diffuse neutrino emission from the Galactic Plane~\cite{IceCube:2023ame} and all-sky~\cite{IceCube:2024fxo}. We consider the most standard scenario of Hawking evaporation of Schwarzschild PBHs directly to emit primary neutrinos and eventually the secondary neutrinos as the decay products from other particles emitted as primaries. The primary Hawking spectrum follows almost a blackbody spectrum given by
\be
\frac{d^{2}N_i^{\rm p}}{dEdt} = \frac{n_{\rm dof}^{i}}{2\pi \hbar}\frac{\Gamma^{i}(M_{\rm PBH},E)}{(e^{E/(k_B T_{\rm PBH})} - (-1)^{2s})} \,,
\ee
where the PBH temperature is $T_{\rm PBH}= \hbar c^3/\big(8 \pi G k_B M_{\rm PBH} \big)$, $k_B$ is the Boltzmann constant, $\Gamma^{i}(M_{\rm PBH},E)$ is the so called greybody factor which depends on the mass (and spin) of the PBH as well as the energy $E$ and spin $s$ of the emitted particle $i$, and $n_{\rm dof}^{i}$ is the number of degrees of freedom of particle $i$. We consider the emission of all three flavors of neutrinos (and antineutrinos), therefore we have\footnote{Note that we assume the neutrinos to be Dirac. For Majorana neutrinos $n_{\rm dof}^\nu = 6$. While Hawking emission produces all helicity states, only left-handed neutrinos and right-handed antineutrinos participate in weak interactions relevant for detection and we account for this consistently.} $n_{\rm dof}^{\nu} = 12$. We extract the greybody factors from \texttt{BlackHawk v2.3}~\cite{Arbey:2019mbc,Arbey:2021mbl} and also use it to compute the secondary neutrino emission. The total neutrino spectra from the evaporation of a single PBH can then be obtained as the sum of the primary and secondary components, that is, $d^2 N_\nu^{\rm tot}/\big(dE_\nu dt \big) = d^2 N_\nu^{\rm p}/\big(dE_\nu dt \big) + d^2 N_\nu^{\rm s}/\big(dE_\nu dt \big)$.

We assume that our Universe is populated by a distribution of PBHs spanning a range of masses. In realistic formation scenarios, an extended mass function at formation $\Psi(M,t=0)$ is expected, with its shape determined by the underlying production mechanism. We adopt a formation scenario based on critical collapse, which naturally gives rise to an extended low mass tail~\cite{Choptuik:1992jv,Niemeyer:1999ak}.
Specifically, we use a parametrization of the mass function at formation that generalizes the standard critical collapse model by allowing independent control of the mass tails, commonly referred to as generalized critical collapse (GCC). This choice is motivated by the results of~\cite{Gow:2020cou}, which show that the GCC mass function provides an excellent fit to the distribution produced by a peak in the primordial power spectrum. The GCC mass function, defined in Appendix~\ref{appsec:mass_func}, is characterized by three parameters: the peak mass $\bar{M}_{\rm PBH}$, and $\alpha$ and $\beta$, which control the slope of the low mass tail and the exponential cutoff at high masses, respectively. An initial PBH population evolves over cosmological timescales to the present epoch ($t=t_0$, the current age of the Universe). During this evolution PBHs lose mass via evaporation, thereby modifying the initial mass function. We account for this time evolution of the mass function as described in Appendix~\ref{appsec:mass_func}.

We separate the contribution to the galactic and the extra-galactic high-energy neutrino flux since we assume different spatial distributions for the PBH population. The galactic contribution from a given direction $\hat{n} = (\ell,b)$ can be computed using
\begin{align}
\phi_\nu^{\rm G} (E_\nu, \ell,b) &= \frac{1}{4\pi} \int_{\rm los} ds\ \rho_{\rm PBH} \big( \mathbf r(s,\ell,b) \big) \int_{M_{\rm PBH}^{\rm min}}^{M_{\rm PBH}^{\rm max}} dM \nonumber \\
&\frac{\Psi(M,t_0)}{M} \frac{d^2 N_\nu^{\rm tot}}{dE_\nu dt}(E_\nu, M)\,, 
\end{align}
where the Galactocentric radius $|\mathbf r(s,\ell,b)| = r =  \sqrt{R_0^2 + s^2 - 2R_0 s \cos \ell \cos b}$ with $\ell$ and $b$ as the Galactic longitude and latitude respectively (so the Galactic Center has $(\ell,b) = (0,0)$), $s$ is the distance along the line of sight (los), $R_0$ is the distance of the sun from the Galactic Center. The density of PBHs at a given position $\mathbf r$ is given by $\rho_{\rm PBH} (\mathbf r) = f_{\rm PBH} \rho_{\rm DM} (\mathbf r)$, where $f_{\rm PBH}$ is the fraction of DM that the PBHs make up, given that the total DM density at that position is $\rho_{\rm DM} (\mathbf r)$. We assume a spherically symmetric DM halo~\cite{Davis:1985rj,1991ApJ...383..112F} with a Navarro-Frenk-White (NFW) profile~\cite{Navarro:1995iw}, such that $\rho_{\rm DM} (\mathbf r) = \rho_{\rm DM} (r) = \rho_s/\big( x (1+x)^2 \big)$, where the dimension-less radius $x = r/r_s$, $r_s$ is the scale radius, and $\rho_s$ is the normalization such that $\rho_s = 4 \rho_{\rm DM} (r = r_s)$. We have $R_0 = 8.2$ kpc, $r_s = 24.42$ kpc, $\rho_s = 0.353\ {\rm GeV\ cm}^{-3}$, such that the local DM density $\rho(R_0) \approx 0.589\ {\rm GeV\ cm}^{-3}$~\cite{Benito:2019ngh}, $M_{\rm PBH}^{\rm min} = 10^{9}$g, $M_{\rm PBH}^{\rm max} = 10^{20}$g, $\alpha = 1$, $\beta = 2.78$ and $t_0 = 13.79$ Gyr. Note that the first integral over the line of sight gives a typical J-factor like column density and has all the directional dependence. The second integral covers the total neutrino emission from all PBHs. For the galactic contribution, in particular, we need the PBH mass function today, that is $\Psi(M,t=t_0)$. Finally, for a given region in the sky with solid angle $\Delta \Omega (= \int d\ell \int db\ \cos b)$ the average neutrino intensity can be defined as
\be
\label{eq:gal_avg}
\langle \phi_\nu^{\rm G} (E_\nu) \rangle_{\Delta \Omega} = \frac{1}{\Delta \Omega} \int_{\ell_{\rm min}}^{\ell_{\rm max}}\ d\ell \int_{b_{\rm min}}^{b_{\rm max}} db\ \cos b\ \phi_\nu^{\rm G} \big( E_\nu, \ell , b \big)\,.
\ee
The average all-flavor galactic neutrino contribution from the evaporation of PBHs with $\bar{M}_{\rm PBH} = 10^{15}$ g with $f_{\rm PBH} = 10^{-4}$ (darker) and $10^{17}$ g with $f_{\rm PBH} = 10^{-2}$ (lighter) is shown in Fig.~\ref{fig:gal_egal_flux} (dashed lines), where we take $\ell_{\rm min} = -180^\circ$, $\ell_{\rm max} = 180^\circ$, $b_{\rm min} = -90^\circ$, and $b_{\rm max} = 90^\circ$.

The extra-galactic component can be defined as
\begin{align}
\label{eq:exgal}
\phi_\nu^{\rm EG} (&E_\nu^{\rm obs}) = \frac{c}{4\pi} \int_0^{z_{\rm max}}dz  \frac{\rho_{\rm PBH}(z)}{(1+z) H(z)} \int_{M_{\rm PBH}^{\rm min}}^{M_{\rm PBH}^{\rm max}} dM \nonumber\\
&\left. \frac{\Psi\big( M,t(z)\big)}{M} \frac{d^2 N_\nu^{\rm tot}}{dE_\nu dt} (E_\nu, M) \right|_{E_\nu = (1+z)E_\nu^{\rm obs}}\,,
\end{align}
where the observed neutrino energy $E_\nu^{\rm obs}$ and the emitted neutrino energy are related as $E_\nu = (1+z)E_\nu^{\rm obs}$, $z$ is the redshift, $t(z) = \int_z^\infty dz'/\big( (1+z') H(z') \big)$, the PBH density is given by $\rho_{\rm PBH}(z) = f_{\rm PBH} \rho_{\rm DM,0} (1+z)^3$ since they contribute a fraction $f_{\rm PBH}$ to the average DM density and therefore scale as non-relativistic matter, and $\rho_{\rm DM,0}$ is the present DM density. We assume a spatially flat $\Lambda$CDM Universe where the Hubble expansion rate is given by $H(z)$. Note that unlike in the case of galactic contribution, for the extra-galactic component we need a time-dependent mass function $\Psi\big( M,t(z) \big)$, since the PBH mass function evolves over cosmic time $t(z)$ due to evaporation, as previously described. We assume $\rho_{\rm DM,0} = 1.26 \times 10^{-6}$ GeV/cm$^3$, $z_{\rm max} = 100$, $H_0 = 67.4$ km/s/Mpc, $\Omega_m = 0.315$, and $\Omega_\Lambda = 0.685$~\cite{Planck:2018vyg}. The results for the extragalactic diffuse flux $\phi_\nu^{\rm EG}$ is shown in Fig.~\ref{fig:gal_egal_flux} (dotted lines). The total contribution from the galactic and the extra-galactic components can be evaulated by $\phi_\nu = \langle \phi_\nu^{\rm G} \rangle_{\Delta \Omega} + \phi_\nu^{\rm EG}$ and is also shown in Fig.~\ref{fig:gal_egal_flux} (solid lines). We see that $\phi_\nu^{\rm EG}$ has a peak that is shifted to lower energies due to contributions from higher redshifts. 

The neutrino flux from the \emph{Galactic Ridge} region ($|\ell| < 30^\circ, |b|<2^\circ$) of the Milky Way has been constrained by observational data from the ANTARES telescope for $E_\nu$ between $1-100$ TeV~\cite{ANTARES:2022izu}. We show the $90\%$ C.L. region of the constraints in Fig.~\ref{fig:gal_egal_flux} as the light brown shaded region. The IceCube Collaboration on the other hand identified the contribution of high-energy neutrinos from the \emph{Galactic Plane} ($|\ell| < 180^\circ, |b|<15^\circ$) at around $4.5 \sigma$ C.L.~\cite{IceCube:2023ame}. The resulting best fitting flux normalizations assuming the $\pi^0$ and KRA$^\gamma_{50}$ models (see also Ref.~\cite{DeLaTorreLuque:2025zsv}) based on the diffuse gamma-ray and cosmic ray observations of the Galactic Plane is shown in Fig.~\ref{fig:gal_egal_flux} with the $90\%$ uncertainty band. The diffuse astrophysical neutrino spectrum based on $10.3$ yrs of IceCube data for the energy range $300$ GeV to $100$ PeV, with improved directional and energy reconstruction was presented in Ref.~\cite{IceCube:2024fxo}. In Fig.~\ref{fig:gal_egal_flux} we show the broken power-law fit to the flux and the $90\%$ uncertainty band corresponding to the single power-law fit. These observational constraints to the high-energy neutrino data allow us to effectively place constraints on $f_{\rm PBH}$ given $\bar{M}_{\rm PBH}$.

We present the results in Fig.~\ref{fig:res} for $\alpha = 1$ and $\beta = 2.78$. The constraints are obtained conservatively by demanding that $\phi_\nu$ from PBH evaporation does not exceed the observational upper limits for \emph{all} $E_\nu$, that is, $\phi_\nu(E_\nu) \leq \phi_{\nu}^{\rm obs,lim}(E_\nu)\ \forall\ E_\nu$, which gives the limit on $f_{\rm PBH}$ as
\be
\label{eq:fpbhlim}
f_{\rm PBH}^{\rm lim}(\bar{M}_{\rm PBH}) = \min_{E_\nu} \left[\frac{\phi_\nu^{\rm obs,lim}(E_\nu)} {\phi_\nu (E_\nu,\bar{M}_{\rm PBH})} \right]\,.
\ee
For the Galactic Ridge and the Galactic Plane results from ANTARES and IceCube respectively we set $\phi_\nu^{\rm EG} = 0$ and compute $\langle \phi_\nu^{\rm G}\rangle_{\Delta \Omega}$ with the appropriate $\ell$ and $b$. The thickness of the bands is given by the corresponding $90\%$ C.L. of the observational constraints. In particular, for IceCube we place our constraints by taking into account both the $\pi^0$ and KRA$^\gamma_{50}$ model templates, where the latter provide a better constraint. For the diffuse flux limits from IceCube we include both the average galactic and extra-galactic components and compare with the broken-power law fit to the data~\cite{IceCube:2024fxo}. It is remarkable to note that all our constraints can rule out $f_{\rm PBH} = 1$ for $\bar{M}_{\rm PBH} \lesssim 3\times 10^{17}$g, with the best constraint reaching up to $\bar{M}_{\rm PBH} \sim 10^{18}$g with existing high-energy neutrino observations. Our bounds are very mildly-dependent on $\beta$ but get weaker by a factor of $\sim 10$ when $\alpha$ is doubled. We also show the existing most constraining monochromatic bounds~\cite{Berteaud:2022tws,Korwar:2023kpy} in the mass range of interest scaled for the GCC mass function following the prescription in~\cite{Carr:2017jsz}. 

Given the reasonable prospects of the current high-energy neutrino detectors, it is interesting to look at the scenario in the era of more powerful upcoming detectors (see Fig.~\ref{fig:res} \emph{right panel}). Given the larger size of IceCube-Gen2 over IceCube and better angular and energy reconstruction techniques, a factor of $\sqrt{5}$ improvement in the sensitivity can be reasonably assumed for a background limited regime. For KM3NeT we assume an improvement in sensitivity by a factor of $3.5$ over the ANTARES limits of the Galactic Ridge. Therefore, in the era of upcoming neutrino detectors, we can hope to rule the possibility of $f_{\rm PBH} = 1$ for $\bar{M}_{\rm PBH} \lesssim 2 \times 10^{18}$g, which not only is closer to the existing bounds but also probes further into the asteroid mass gap.

Following Ref.~\cite{Boluna:2023jlo} (see Eq. 3.8 there), we also checked that for the GCC mass function with $\alpha=1$ and $\beta=2.78$, the maximum local burst rate allowed by the current IceCube Galactic plane diffuse flux bounds is $\dot{n}^{\rm max}_{\rm PBH} \lesssim 0.4-2 \text{ pc}^{-3}\text{yr}^{-1}$ over the range of $\bar{M}_{\rm PBH}$ considered, which is $\mathcal{O}(10^3)$ less than the value inferred by assuming some of the IceCube diffuse flux events to originate from PBHs in Ref.~\cite{Klipfel:2025jql}.

\textit{\textbf{Constraints from singular PBH transits.}}
\begin{figure}
\centering
\includegraphics[width=0.48\textwidth]{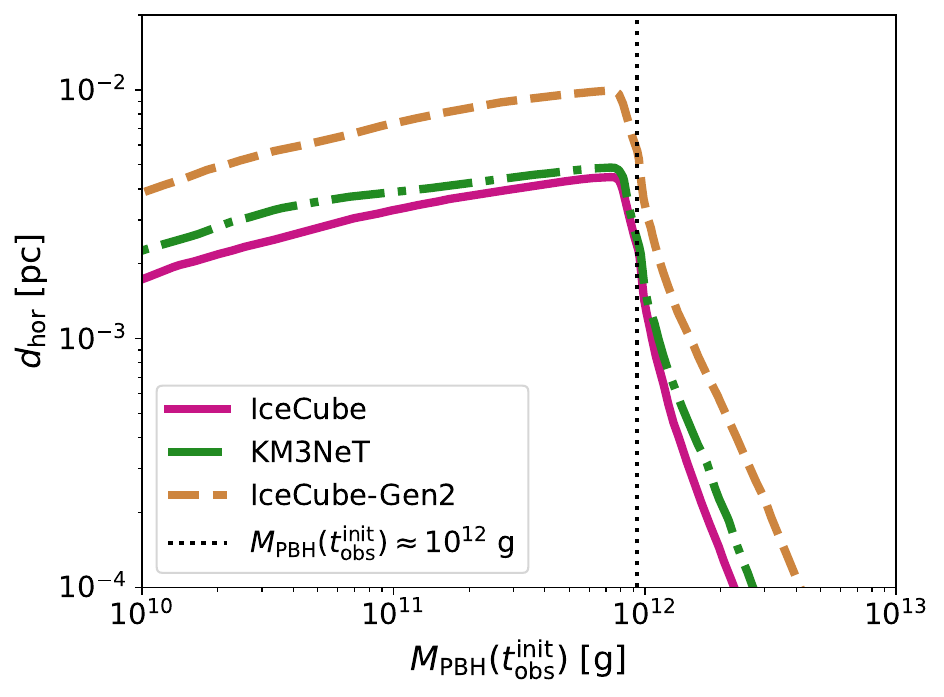}
\caption{\label{fig:dhor}The maximum horizon distance ($d_{\rm hor}$) for IceCube, IceCube-Gen2, and KM3NeT assuming $12.5$ years of operation time for a $5\sigma$ observation of PBH transit, shown as a function of the mass of the PBH at the start of the observation $t_{\rm obs}^{\rm init}$, that is, $M_{\rm PBH}(t_{\rm obs}^{\rm init})$. The vertical line corresponds to $M_{\rm PBH} = 9.3 \times 10^{11}$ g, corresponding to a remaining PBH lifetime of 12.5 yr, which is the maximum time of observation considered.
}
\end{figure}
\begin{figure}
\centering
\includegraphics[width=0.45\textwidth]{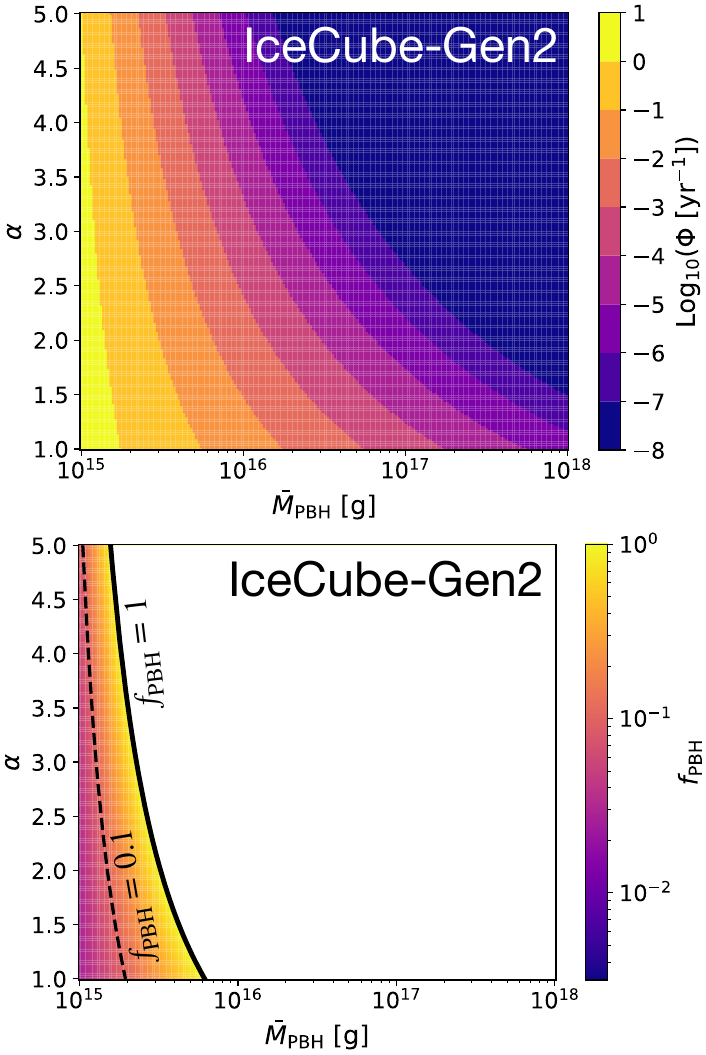}
\caption{\label{fig:trans_fpbhlim}\emph{Top: }Color map of transit rate of PBHs for IceCube-Gen2 in the $\alpha - \bar{M}_{\rm PBH}$ plane given the detectors horizon distance assuming a $5\sigma$ C.L. for a background free scenario. \emph{Bottom: }Color map of $f_{\rm PBH}$ given the transit rate and assuming $12.5$ years of operation on the same plane. We set $\beta = 2.78$ when we vary $\alpha$.
}
\end{figure}
Given the large sky coverage of high-energy neutrino detectors, one may ask whether they can resolve the transit of a single PBH. Since these telescopes are most sensitive at energies above a few hundred GeV, only the lightest and hottest PBHs in the tail of the mass function today can emit a detectable neutrino flux. On the other hand, their large effective areas and wide field of view (FOV) make them potentially sensitive to PBH fly-bys at reasonably large distances (fraction of a parsec). 

We assume a simple fly-by geometry in which the PBH moves at a constant velocity $v_{\rm PBH} = 200\,\text{km/s}$, a characteristic DM velocity (we also tried other values up to 300 km/s and the results do not change significantly), and travels parallel to the detector such that the closest point between the PBH and the neutrino telescope $d_{\rm hor}$, i.e. the impact parameter, corresponds to the exact center of the PBH trajectory, see Appendix~\ref{appsec:ic_km3_trans_fpbhlim} for a simple sketch of the geometry considered. We then compute the maximum $d_{\rm hor}$ for which a given neutrino telescope can achieve a $5\sigma$ detection over a maximum observation window of $t_{\rm obs} = 12.5$ yr. The significance is defined as $ S = N_\nu/\big(\sqrt{N_\nu+N_{\rm bckg}}\big)$, where $N_{\rm bckg}$ is the number of background neutrino events and $N_\nu$ is the number of events from the PBH, given by
\ba
N_\nu (d_{\rm hor}) = \frac{1}{4 \pi} &&\int_{E_\nu^{\rm min}}^{E_\nu^{\max}} dE_\nu A_{\rm eff}(E_\nu) \times\nonumber\\
&&\int_{t^{\rm init}_{\rm obs}}^{t^{\rm fin}_{\rm obs}} 
dt \frac{1}{d(t)^2} \frac{d^2 N_\nu^{\rm tot}}{dE_\nu dt}\big(M_{\rm PBH}(t) \big)\,,
\ea
where $d(t)=\sqrt{d_{\rm hor}^2+(D/2-v_{\rm PBH} t)^2}$. The PBH travels a distance $D = v_{\rm PBH} \Delta t_{\rm obs}$, where $\Delta t_{\rm obs} = t^{\rm fin}_{\rm obs} - t^{\rm init}_{\rm obs}$ is determined either by the PBH remaining lifetime or by the maximum observation window of $12.5$ yrs. We include both primary and secondary neutrino emission and integrate over the energy range $E_\nu^{\rm min} = 1$ GeV to $E_\nu^{\rm max} = 10^6$ GeV, although below a few hundred GeV the effective areas $A_{\rm eff}$ of the telescopes considered become very small, rendering the contribution from lower energies negligible. The effective areas for IceCube including DeepCore are estimated from Refs.~\cite{IceCube:2023yfi,IceCube:2025gkd,IceCube:2025lev}, while that for KM3NeT including ARCA and ORCA from Ref.~\cite{KM3NeT:2024paj,KM3NeT:2025mkg,Bozza:2023slv}, and for IceCube-Gen2 we scale $A_{\rm eff}$ by a factor of $10^{2/3}$~\cite{IceCube-Gen2:2020qha,Gen2_TDR}. 

We also set $N_{\rm bckg} = 0$ and the effects of background are discussed in Appendix~\ref{appsec:bkg}. 
The absence of backgrounds can be justified by the following reasons. First, instead of using the all-sky effective area for our computations for IceCube, we only consider the Northern Sky, where the Earth acts as a shield, significantly reducing the atmospheric muons and neutrinos. IceCube identifies up-going tracks and provides a much cleaner sample of events that can be used for dedicated searches. Additionally, joint observations with gamma-ray telescopes like HAWC~\cite{Abeysekara:2017hyn,Abeysekara:2017mjj} and LHASSO~\cite{LHAASO:2019qtb,LHAASO:2021crt} which have sub-degree angular resolution, will allow the remaining isotropic backgrounds to be heavily reduced by limiting searches to the localized sky patches. We show the results for the maximum $d_{\rm hor}$ in Fig.~\ref{fig:dhor} for IceCube, IceCube-Gen2, and KM3NeT.

For PBH masses $M_{\rm PBH} \lesssim 10^{12}\,\text{g}$, the detector observes the PBH over its entire remaining lifetime. For larger masses, the observation is limited to a period of 12.5 yr, during which the PBH mass remains approximately constant. In this latter case, the detector does not observe the final exploding phase, when the PBH emits higher-energy neutrinos in the range $E_\nu \sim 10^3$–$10^6$ GeV, and consequently $d_{\rm hor}$ decreases significantly. Increasing the maximum observation time could, in principle, shift this drop toward larger PBH masses. However, since the PBH lifetime scales as $\tau_{\rm PBH} \propto M_{\rm PBH}^3$, achieving sensitivity to higher masses would require extending the observation time by several orders of magnitude beyond 12.5 yr, which is not ideal.

We then compute the transit rate following the prescription outlined in Appendix~\ref{appsec:ic_km3_trans_fpbhlim}. As before, we assume a GCC mas function and scan over the parameters $\alpha$, $\beta$ and $\bar{M}_{\rm PBH}$. In Fig.~\ref{fig:trans_fpbhlim} \emph{(top)} we show the results for a fixed value of $\beta$ for IceCube-Gen2 (we show the results for other telescopes in Appendix~\ref{appsec:ic_km3_trans_fpbhlim}). We also checked that the results are very mildly dependent on $\beta$ for a fixed $\alpha$, since this is the parameter that controls the exponential fall-off at higher PBH masses. As we can see, only when the mass function is peaked at lower values $\bar{M}_{\rm PBH} \lesssim 10^{16}$g, we expect one singular transit over 1-10 yr(s) of observation. We note that these results are comparable to the expected performance of the enhanced AMS detector with lower positron threshold of 50 MeV proposed in Ref.~\cite{Klipfel:2025bvh}.

We can now derive upper bounds on the PBH abundance using 12.5 years of IceCube observations (we assume the same observation window for IceCube-Gen2 and KM3NeT) and the absence of confirmed PBH transit events. Although several neutrino events from unidentified point sources have been reported~\cite{Klipfel:2025jql}, there is currently no evidence that the PBH hypothesis is statistically favored; we therefore do not interpret these events as PBH signals. The resulting constraints are shown in Fig.~\ref{fig:trans_fpbhlim} \emph{(bottom)}. As is evident from the figure, these bounds are significantly weaker than those derived from the diffuse neutrino flux and other existing constraints. Despite their large effective areas and sensitivity to PBH transits at sub-parsec distances, high-energy neutrino telescopes primarily probe the tail of the mass function, leading to a strongly suppressed transit rate. The corresponding limits on $f_{\rm PBH}$ for current (\emph{left panel}, dotted curve) and future (\emph{right panel}, dot-dashed curves) neutrino telescopes, assuming $\alpha = 1$ and $\beta = 2.78$, are also shown in Fig.~\ref{fig:res}.

\textit{\textbf{Conclusions.}}
Motivated by recent advances in high-energy neutrino astronomy, the extensive data set accumulated over more than a decade of observations with IceCube, the KM3NeT $220$ PeV event, and the ambitious experimental programs of current and next-generation neutrino telescopes (KM3NeT and IceCube-Gen2), we for the first time investigate the role of high-energy neutrinos, serving as an independent messenger, in constraining the abundance of PBHs. In particular, we consider both the diffuse neutrino background arising from PBH populations in the Galaxy and across cosmological distances, as well as the potential detection of an individual PBH during a close fly-by. Since we focus on high-energy neutrino observations, only relatively light PBHs ($M_{\rm PBH}\lesssim 10^{12}$g today) can produce an observable signal in the energy range of interest. Nevertheless, realistic formation scenarios generically predict extended PBH mass functions that can populate this range of masses. In this work, we adopt as a benchmark the critical collapse scenario, based on the collapse of large overdensities upon horizon reentry, which naturally yields an extended distribution with a pronounced low mass tail. 

Crucially, although high-energy neutrino astronomy restricts sensitivity to a limited PBH mass range, the large effective areas and wide FOV of high-energy neutrino telescopes provide a significant advantage over other probes, partially compensating for this limitation. For a well-motivated yet optimistic choice of mass function parameters, we exclude the possibility that PBHs constitute the entirety of DM for $M_{\rm PBH} \lesssim 10^{18}\,\mathrm{g}$ using existing data, and up to $M_{\rm PBH} \lesssim 2 \times 10^{18}\,\mathrm{g}$ with next-generation detectors  (see Fig.~\ref{fig:res}). These limits lie within an order of magnitude of current leading constraints, underscoring the competitiveness of neutrino-based probes. Furthermore, consistent with earlier indications in the literature~\cite{Airoldi:2025opo}, our analysis of single PBH transit events shows that their detection with high-energy neutrino telescopes is exceedingly unlikely given existing bounds.

Our limits on $f_{\rm PBH}$ from the diffuse neutrino flux are conservative and these can be improved by considering the signal template and performing $\chi^2$ likelihood estimates, which we leave for future work. Furthermore, we do not subtract\footnote{Assuming some dominant contributors to the diffuse flux like AGN, blazars, supernovae, and then subtracting the corresponding fluxes from the diffuse flux, would also make our limits stronger.} any source contributions from the diffuse flux.
We also assume solely Schwarzschild PBHs which emit just SM particles. Going beyond this assumption can significantly modify the PBH evolution and consequently the corresponding bounds~\cite{Baker:2025zxm}. Larger values of spin could also result from dynamical evolution processes (see for instance~\cite{Berti:2008af}), leading to the formation of Kerr black holes, which also yield stronger bounds (see Appendix~\ref{appsec:kerr_pbhs}). Finally, existing low-energy neutrino limits from Super-K, and projected ones for Hyper-K, JUNO, and DUNE~\cite{Dasgupta:2019cae,Bernal:2022swt} are compared to the high-energy limits we obtain in this work in Appendix~\ref{appsec:nu_lim} (albeit for the log-normal mass function) where we find that the latter are more constraining in the relevant mass range.

Our work is particularly timely because inevitably with more precise measurements of the Galactic plane contribution to the diffuse neutrino flux from IceCube or more data from KM3NeT, our constraints can be significantly improved. In that case, such constraints besides using an independent messenger to diminish the parameter space of $f_{\rm PBH}$, can also provide leading constraints in the asteroid mass gap. Our current work serves as a step in highlighting this aspect of high-energy neutrinos and corresponding detectors in the context of whether PBHs can be a significant fraction of DM.
%
\acknowledgments
\textit{\textbf{Acknowledgements.}}
We thank Yuber F. Perez-Gonzalez, Pedro De La Torre Luque, and Sarunas Verner for useful discussions and comments.
M.\,M. acknowledges support from the FermiForward Discovery Group, LLC under Contract No. 89243024CSC000002 with the U.S. Department of Energy, Office of Science, Office of High Energy Physics.
\bibstyle{apsrev4-2}
\bibliography{refs}
\clearpage
\newpage
\appendix
\onecolumngrid
\maketitle
\onecolumngrid
\begin{center}
	\textbf{\Large Supplementary Material}
	 \bigskip\\
		\textbf{\large High-energy neutrino constraints on primordial black holes as dark matter}
		 \medskip\\
   {Mainak Mukhopadhyay and Joaquim Iguaz Juan}
\end{center}
\renewcommand{\thesection}{S\arabic{section}}
\renewcommand{\theequation}{S\arabic{equation}}
\renewcommand{\thefigure}{S\arabic{figure}}
\renewcommand{\thetable}{S\arabic{table}}
\renewcommand{\thepage}{S\arabic{page}}
\setcounter{equation}{0}
\setcounter{figure}{0}
\setcounter{table}{0}
\setcounter{page}{1}
\setcounter{secnumdepth}{4}
\section{On mass functions and their evolution}
\label{appsec:mass_func}
In this section, we elaborate on the extended mass functions, in particular the generalized critical collapse (GCC) and log-normal (LN) mass functions. The former is what we primarily used all throughout this work, while the latter will be relevant in comparing our results with previous work on low energy neutrino bounds in Appendix~\ref{appsec:nu_lim}. Assuming the PBHs form in the early universe at time $t_{\rm init}$ with masses $M_i$, the initial comoving number density of PBHs in between $[M_i, M_i+dM_i]$ is defined as $d n_i/dM_i$. Therefore the total initial PBH mass density can be computed as $\rho_{{\rm PBH},i} = \int dM_i\ M_i(dn_{{\rm PBH},i}/dM_i)$. This initial comoving number density can be used to define an initial normalized differential mass density function as 
\be
\Psi (M_i,t_i) \equiv \Psi_i = \frac{1}{\rho_{{\rm PBH},i}} \frac{d\rho_{{\rm PBH},i}}{dM_i} = \frac{M_i}{\rho_{{\rm PBH},i}}\frac{d n_{{\rm PBH},i}}{dM_i}\,,\ {\rm and\ }\int dM_i\ \Psi_i = 1\,,
\ee
where, the quantity $\Psi_i$ is known as the initial mass function.

As mentioned earlier, PBHs can be formed in a process known as critical collapse~\cite{Choptuik:1992jv,Niemeyer:1999ak}, that results in a mass function with an extended low mass tail. We parameterized the mass function using the GCC parameterization~\cite{Gow:2020cou} as follows
\be
\Psi_i^{\rm GCC} (M_i|\mu, \alpha,\beta) = \frac{\beta}{\mu} \frac{1}{\Gamma\big( (\alpha+1)/\beta \big)} \bigg( \frac{M_i}{\mu} \bigg)^\alpha \exp \bigg( -\big( M_i/\mu \big)^\beta \bigg)\,,{\ \rm and}\ \bar{M}_{\rm PBH} (\mu,\alpha,\beta) = \mu \big( \alpha/\beta \big)^{1/\beta}\,,
\ee
where the function peaks at $\bar{M}_{\rm PBH}$ and the low and high mass tails are governed by $\alpha$ and $\beta$ respectively.

The normalized log-normal mass function can be similarly parameterized as
\be
\Psi_i^{\rm LN} (M_i,\mu,\sigma) = \frac{1}{\sqrt{2 \pi} \sigma M_i} \exp \bigg( -\frac{\log \big( M_i/\mu \big)^2}{2\sigma^2} \bigg)\,,{\ \rm and}\ \bar{M}_{\rm PBH} (\mu,\sigma) = \mu\ \exp \big( -\sigma^2 \big)\,,
\ee
where again $\bar{M}_{\rm PBH}$ which sets the location of the peak and $\sigma$ which governs the width of the mass function.

The initial PBH mass distribution can change over cosmological timescales due to processes like Hawking evaporation, PBH mergers, or accretion. In the range of masses we consider in this work ($M_{\rm PBH}<10^{18}$g), the two later processes can be safely neglected. The time evolution of the mass function will then be given by the rate of PBH evaporation of different masses. The PBH mass evolution is governed by~\cite{Page:1976ki,Page:1977um}
\be
\frac{dM}{dt} = -\sum_i g_i \int_{m_i}^\infty E dE \frac{d^2 N_i^{\rm p}}{dE dt} = - \frac{\alpha(M)}{M^2}\,,
\ee
where the Page function $\alpha(M)$ is given by
\be
\alpha(M) \equiv M^2 \sum_i g_i \int_{m_i}^\infty E dE \frac{d^2 N_i^{\rm p}}{dE dt},
\ee
and where the summation runs over all species present in nature (in this work we assume the Standard Model), the quantum degrees of freedom associated with a particle species $i$ is given by $g_i$ and their mass is denoted by $m_i$, and $d^2 N_i^{\rm p}/(dE dt)$ is the primary Hawking spectrum. Solving this equation leads to the relationship between the initial PBH mass $M_{\rm init}$ at the time formation and its mass $M$ at any given instant $t$. Assuming the age of the universe $t_0 \simeq 13.8$ Gyr we find that PBHs that had $M_{\rm init} \gtrsim 5 \times 10^{14}$g survive until today while PBHs with lower masses have already evaporated.

\begin{figure}
\centering
\includegraphics[width=0.8\textwidth]{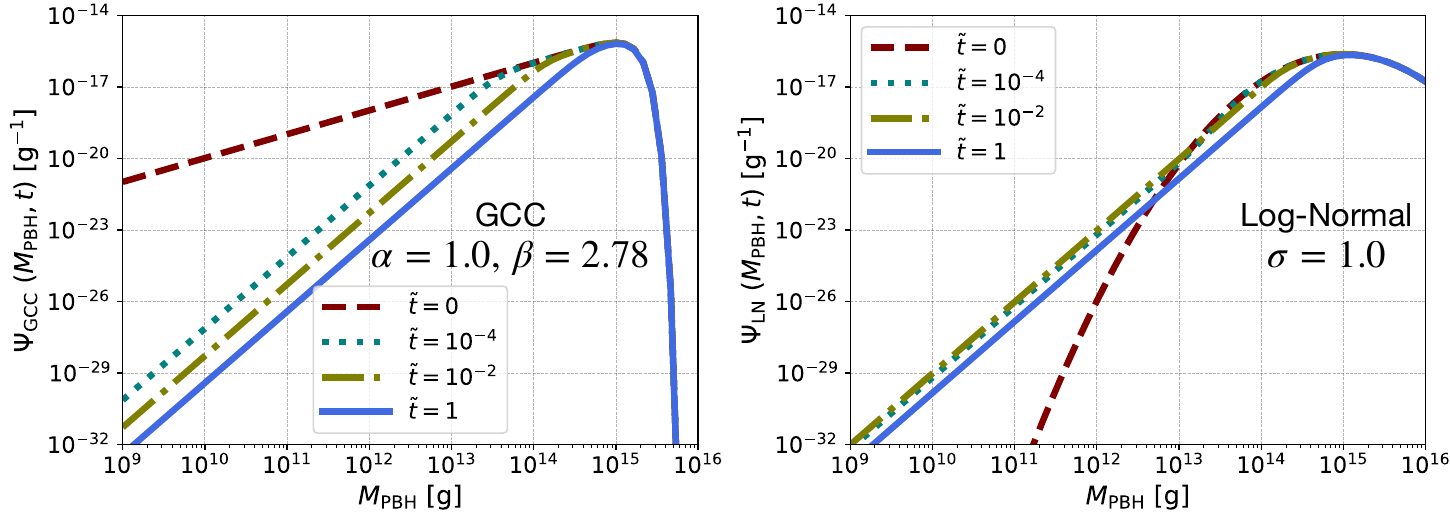}
\caption{\label{fig:massfunc}Time evolution of the GCC mass function ($\Psi_{\rm GCC}$) with $\alpha = 1$, $\beta = 2.78$, and $\bar{M}_{\rm PBH} = 10^{15}$g \emph{(left)} and the log normal mass function ($\Psi_{\rm LN}$) with $\sigma = 1$ and $\bar{M}_{\rm PBH} = 10^{15}$g \emph{(right)}. The time is shown as $\tilde{t} = t/t_0$, where $t_0 \approx 1.3787 \times 10^{10}$ years is the current age of the universe. The solid line ($\tilde{t} = 1$) denotes the current age of the universe while the dashed line ($\tilde{t} = 0$) denotes the initial mass function.
}
\end{figure}

To account for the time evolution of the mass function, we proceed as follows. The normalized PBH number distribution function $\Phi(M_i,t_i) \equiv \Phi_i = (1/n_{{\rm PBH},i}) \big( dn_{{\rm PBH},i}/dM_i \big)$ such that $\int dM_i\ \Phi_i = 1$. One can show using the above relations that $\Phi_i = \mathcal{N} \Psi_i/M_i$, where $\mathcal{N} = \rho_{{\rm PBH},i}/n_{{\rm PBH},i} = 1/\big( \int_0^\infty dM_i (\Psi_i/M_i)\big)$. Now, assume a population of PBHs with masses $M_i$ at the time of formation $t_i$, at given instant in time $t$ they would have masses $M$ such that the number of PBHs in a mass bin $[M_1, M_2]$ is conserved, that is, $n_{{\rm PBH},i} \Phi(M,t) dM = n_{{\rm PBH},i} \Phi_i dM_i$. This readily leads to an expression that relates $\Phi$ to $\Phi_i$, $\Phi(M,t) = \Phi_i \big( M_i (M,t),t_i \big) \big( dM_i(M,t)/dM \big)$. In addition, at any given time $t$ the comoving number density of PBHs in the same mass bin can be defined as 
\be
n_{[M_1,M_2]}(t) = n_{{\rm PBH},i} \int_{M_1}^{M_2} dM\ \Phi(M,t) = n_{{\rm PBH},i} \int_{M_1}^{M_2} dM\ \Phi_i \big( M_i(M,t),t_i \big) \left(\frac{dM_i}{dM} \right) = n_{{\rm PBH},i} \int_{M_i(M_1,t)}^{M_i(M_2,t)} dM_i\ \Phi_i\,.
\ee
From the above equations, we can clearly establish a relation between $\Psi_i$ and $\Psi (M,t)$ given the fact that $\Psi$ and $\Phi$ are related. This can be computed as
\be
\Psi(M,t) = \frac{M}{M_i(M,t)} \Psi \big(M_i(M,t), t_i \big) \frac{dM_i(M,t)}{dM}\,.
\ee
The time-dependent mass function can also be approximately given by the following (see~\cite{Klipfel:2025bvh} for details)
\be
\Psi (M,t) \approx \frac{M^3}{M_i(M,t)} \Psi\big( M_i(M,t), t_i \big) \big( M^3 + 3 t \alpha(M_i) \big)^{-2/3}\,.
\ee
In this work, we use the analytical approximations for computing the mass functions which are reasonably accurate for all-purposes. We fix $\alpha(M) = 1.02\times 10^{26}\ \text{g}^3\text{s}^{-1}$ (this corresponds to fixing $f(M)=1.97$ as in Ref.~\cite{Klipfel:2025bvh}) and thus use the Page factor in a mass-independent way. The results for the time evolution of the mass functions are shown in Fig.~\ref{fig:massfunc}. The GCC mass function has a typical low mass tail at the initial formation time ($\tilde{t} = 0$, where $\tilde{t} = t/t_0$). As time evolves, this low mass tail rapidly drops since the PBHs evaporate leading to a decrease in the number density for $M_{\rm PBH} \ll \bar{M}_{\rm PBH}$. For the log normal case, the mass function is symmetric around the peak to begin with. However, the high mass tail hardly evaporates, while the PBHs in the low mass tail evaporate quickly becoming even lighter. This results in a pile-up and the low mass tail grows with time as compared to the initial time. Interestingly, as already noted in previous works~\cite{DeRomeri:2025dwm}, the current day mass function is mostly independent of the initial mass function as can be seen by comparing the solid light blue curve in the left and right panels of the figure.
\section{Constraints from transit Rates for IceCube and KM3NeT}
\label{appsec:ic_km3_trans_fpbhlim}
\begin{figure}
\centering
\includegraphics[width=0.3\textwidth]{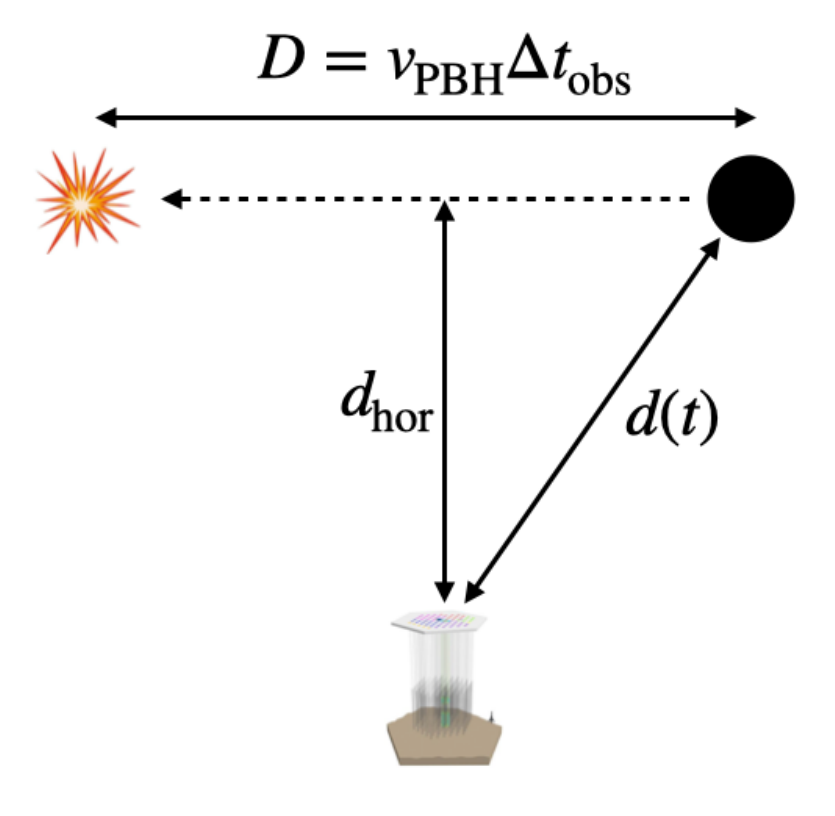}
\caption{\label{fig:geo} Geometry of the PBH fly-by. We assume the PBH trajectory to be always symmetric with respect to IceCube, with $d_{\rm hor}$ the point of closer distance.
}
\end{figure}

In this Appendix we outline the method used to compute the transit rate $\Phi_{\rm transit}$. We assume the simplified geometry shown in Fig.~\ref{fig:geo}, in which the PBH trajectory is symmetric with respect to the neutrino detector. 
\begin{figure}
\centering
\includegraphics[width=0.8\textwidth]{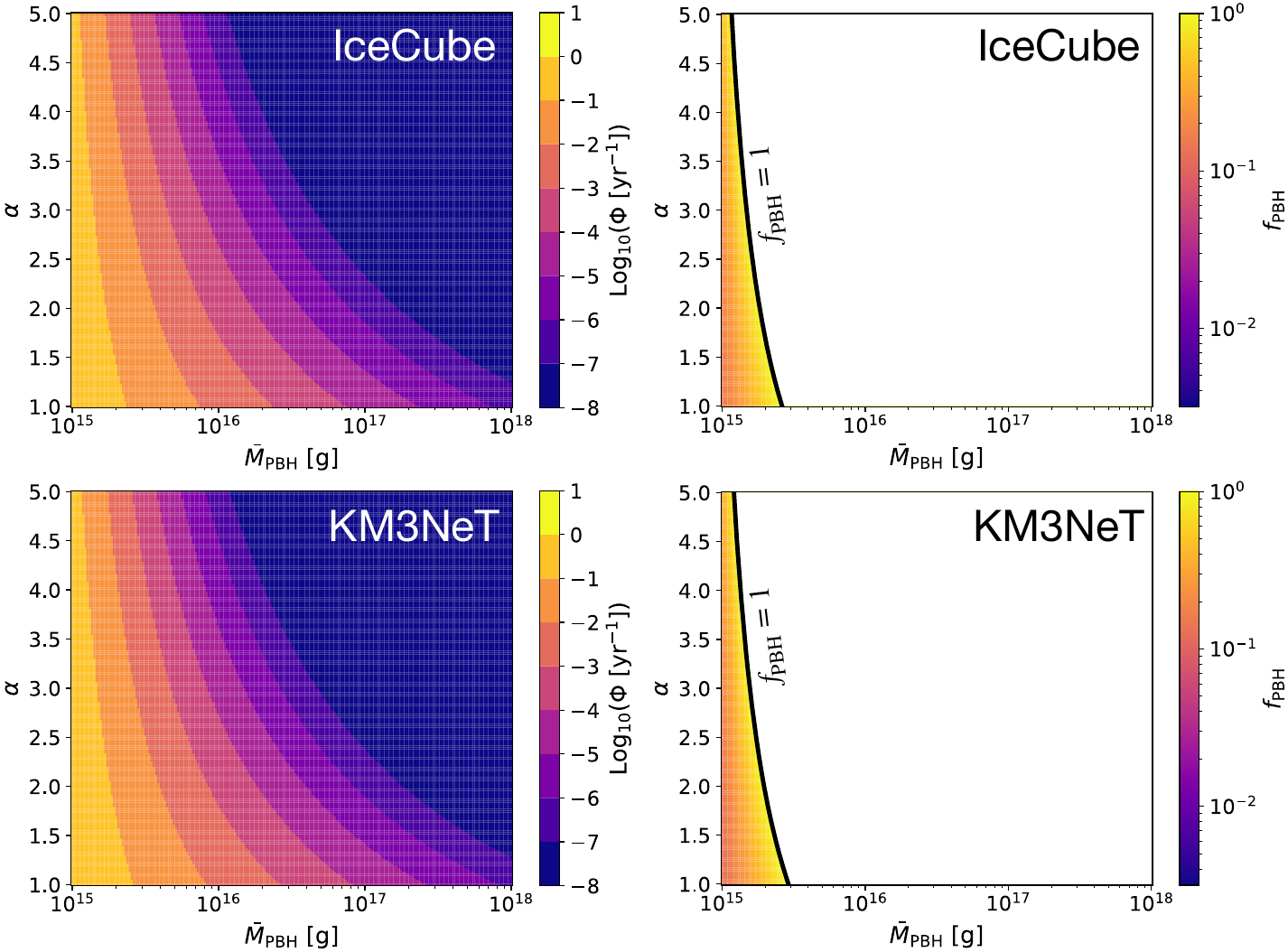}
\caption{\label{fig:trans_IC_km3net}Same as Fig.~\ref{fig:trans_fpbhlim} but for IceCube \emph{(top)} and KM3NeT \emph{(bottom)}.
}
\end{figure}
This allows us to determine the maximum distance $d_{\rm hor}$ at which a given telescope can achieve a $5\sigma$ detection, as described in the main text. We then follow the prescription of~\cite{Klipfel:2025bvh} to compute the transit rate
\be
\Phi_{\rm transit} = \rho_{DM}^{\odot} \int^{M_{\rm max}}_0 dM \frac{1}{M} \Psi (M,t_0) \int^{v_{\rm esc}}_0 dv\ v f(v) \int^{d_{\rm hor}^{\rm max}}_0 db\ 2\pi b,
\ee
where $\rho_{DM}^{\odot}=0.589 \text{ GeV}\text{cm}^{-3}$ is the local DM density~\cite{Benito:2019ngh}. We average over the PBH velocity distribution $f(v)$, assuming it to be Maxwellian and truncated at the escape velocity $v_{\rm esc}=544$ km/s~\cite{Klipfel:2025bvh}. The mass integral is taken up to $M_{\max}=10^{13}$ g, for which we find $d_{\rm hor}^{\rm max} \sim 1$ AU for all telescopes considered in the background-free case. For larger masses (corresponding to smaller $d_{\rm hor}^{\rm max}$), this geometric approximation breaks down, and additional effects such as the Earth's motion around the Sun should be included. Nevertheless, the resulting transit rate depends only weakly on $M_{\max}$, provided $M_{\max} \gtrsim 10^{12}$ g, as discussed in the main text. In Fig.~\ref{fig:trans_fpbhlim} we show the resulting transit rates for the upcoming telescope IceCube-Gen2 while the results for IceCube \emph{(top panels)} and KM3NeT \emph{(bottom panels)} are shown in Fig.~\ref{fig:trans_IC_km3net}, respectively. We note that number of detectable PBH transits per year with current high-energy neutrino telescopes is comparable to the one from AMS~\cite{Klipfel:2025bvh}.
\section{Effects of background on detecting PBH transits}
\label{appsec:bkg}
\begin{figure}
\centering
\includegraphics[width=0.8\textwidth]{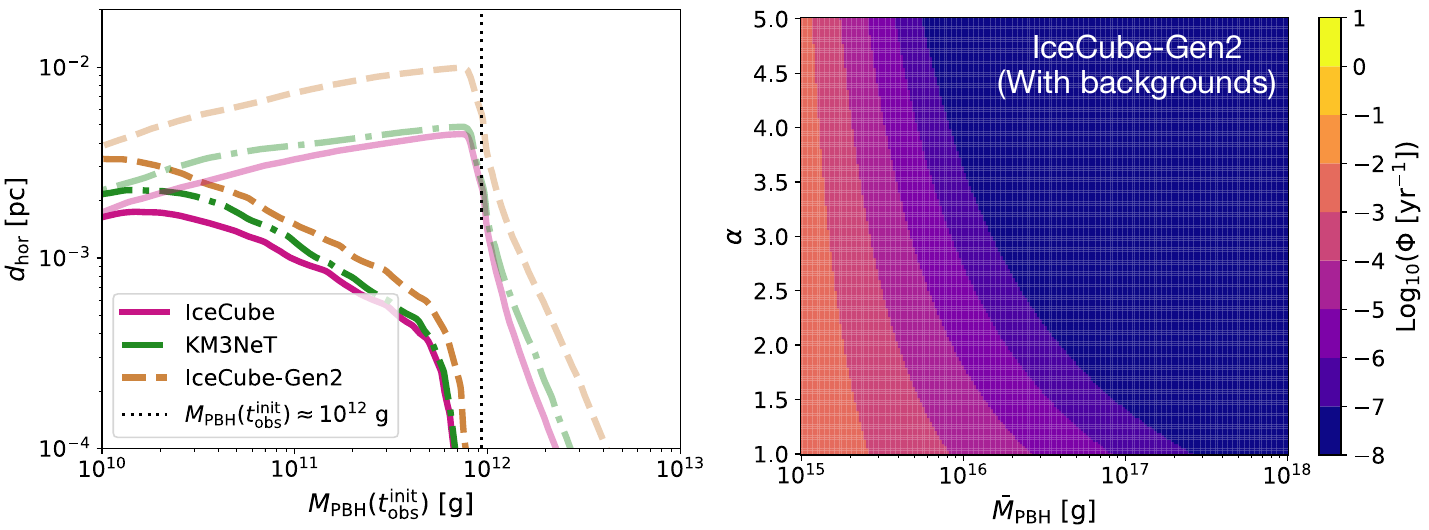}
\caption{\label{fig:dhor_transrate_bkg}\emph{Left: }Same as Fig.~\ref{fig:dhor} but with backgrounds. The results of Fig.~\ref{fig:dhor} without backgrounds are shown in lighter shades. \emph{Right: }Color map of transit rate of PBHs for IceCube-Gen2 in the $\alpha - \bar{M}_{\rm PBH}$ plane given the detectors horizon distance assuming a $5\sigma$ C.L. including backgrounds.
}
\end{figure}
In the main text we neglected backgrounds for our computation of the transit rates. We justified our assumption by considering only Northern Sky events for an IceCube like detector, where the upgoing events can be distinctly separated from the atmospheric muons and neutrinos due to the Earth acting as a shield. Therefore, such a selection criteria leads to a cleaner sample of neutrino events. In this section, we relax that assumption and present the results of the transit rate computation in the presence of backgrounds.

The main high-energy neutrino backgrounds are from the conventional and prompt atmospheric neutrinos and the diffuse astrophysical neutrinos. We include these components in our estimate for backgrounds and neglect the atmospheric muons which would be negligible given our focus on the Northern Sky (for IceCube-like detectors). For neutrino energies $E_\nu \lesssim 10^3$ GeV we use the atmospheric neutrino fluxes computed in Ref.~\cite{Barr:2004br} ($\phi_\nu^{\rm bkg,sub-TeV}$). While for $E_\nu>10^3$ GeV, we estimate the contribution of the conventional atmospheric neutrinos ($\phi_\nu^{\rm bkg,conv.}$) from Ref.~\cite{Honda:2006qj}, while the prompt flux ($\phi_\nu^{\rm bkg,prompt}$) is taken from Ref.~\cite{Enberg:2008te}. The diffuse isotropic astrophysical neutrinos is parameterized using the broken power law (BPL)~\cite{IceCube:2024fxo}, where $\phi_\nu^{\rm bkg, astro} (E_\nu) = \phi_0 \big( E_{\rm break}/100\ {\rm TeV}\big)^{-\gamma_2} \big( E_\nu/E_{\rm break}\big)^{-\gamma_2}$, where the all-flavor normalization $\phi_0 = 5.1 \times 10^{-17}\ {\rm GeV^{-1}cm^{-2}s^{-1}sr^{-1}}$, the spectral index $\gamma_2 = 2.52$, and the energy break $E_{\rm break} = 22.9$ TeV. Therefore, the total background contribution can be estimated as $\phi_\nu^{\rm bkg,tot} (E_\nu) = \phi_\nu^{\rm bkg,sub-TeV} (E_\nu) + \phi_\nu^{\rm bkg,conv.}(E_\nu) + \phi_\nu^{\rm bkg,prompt}(E_\nu) + \phi_\nu^{\rm bkg, astro}(E_\nu)$. In the energy regimes of interest to us, the atmospheric neutrino flux ($\phi_\nu^{\rm bkg,sub-TeV} + \phi_\nu^{\rm bkg,conv.} + \phi_\nu^{\rm bkg,prompt}$) completely dominates over the astrophysical flux, where the latter only becomes relevant for $E_\nu \gtrsim {\rm a\ few\ } 10^5$ GeV.

In Fig.~\ref{fig:dhor_transrate_bkg} we show the effects of including backgrounds. In the \emph{left} panel we show the distance horizon for the various neutrino detectors that enable a $5\sigma$ C.L detection of transiting PBHs, while the \emph{right} panel shows the transit rate for IceCube-Gen2. Similar to Fig.~\ref{fig:trans_fpbhlim} the transit rates are shown in the $\alpha-\bar{M}_{\rm PBH}$ plane corresponding to a distance horizon that enables a $5\sigma$ C.L detection. Clearly, the inclusion of backgrounds reduces the transit rates since now the neutrino events from the PBH transit need to be identified over the background events. However, there are two aspects that are important to be noted here. First, our results are conservative since they assume $5\sigma$ C.L. and second, the background estimate does not include any information from gamma-ray observations. A joint detection strategy with gamma-ray telescopes like HAWC~\cite{Abeysekara:2017hyn,Abeysekara:2017mjj} and LHASSO~\cite{LHAASO:2019qtb,LHAASO:2021crt} will greatly improve the results shown here. This is because with the input from such telescopes, the localization information will be invaluable. With typical sub-degree level source localization, the isotropic backgrounds can be heavily reduced by searching for neutrino events within the small localization area.
\section{Limits for Kerr (spinning) PBHs}
\label{appsec:kerr_pbhs}
\begin{figure}
\centering
\includegraphics[width=0.95\textwidth]{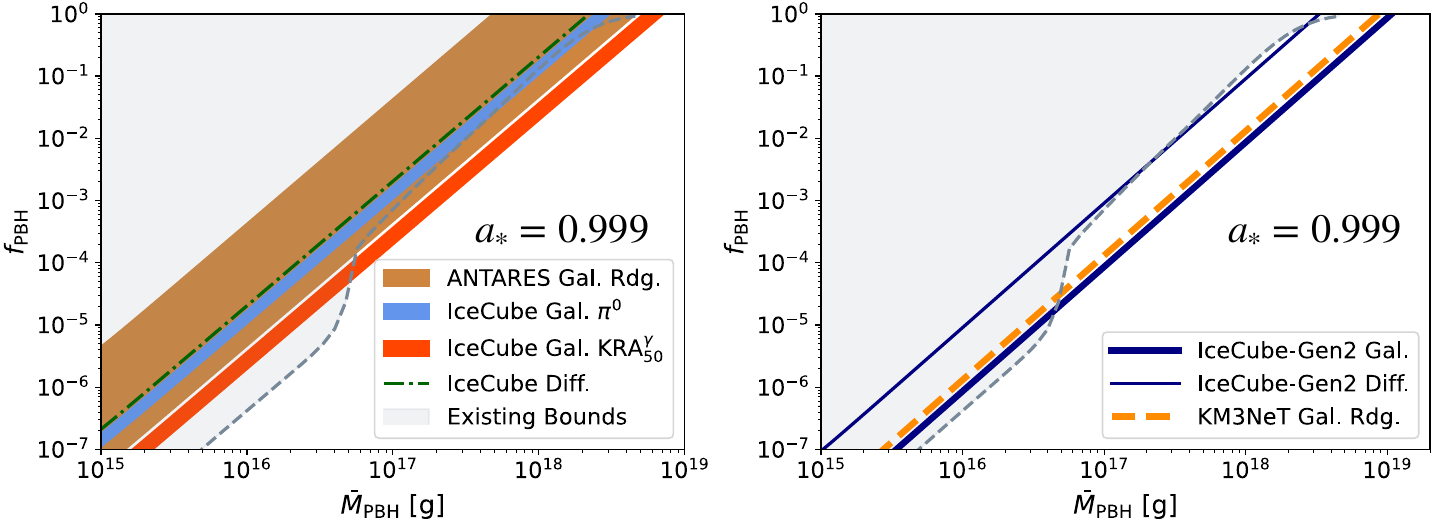}
\caption{\label{fig:spin_fpbhlim}Same as Fig.~\ref{fig:res} but for Kerr PBHs with spin parameter $a_* = 0.999$.  
}
\end{figure}
In some formation mechanisms, PBHs can be born with small spins~\cite{Mirbabayi:2019uph,DeLuca:2019buf}. Larger spin values can subsequently arise from dynamical evolution processes~\cite{Berti:2008af}. Spinning black holes, also known as Kerr black holes, are characterized by their mass and spin parameter $a_* = J /(G M_{\rm PBH}^2)$, where $J$ is the angular momentum of the PBH. Kerr PBHs exhibit enhanced evaporation rates for non-zero spin particles~\cite{Page:1976ki}, in particular neutrinos. It is therefore important to study the dependence of the bounds on the PBH spin.
We compare our limits for Schwarzschild PBHs with those for quasi-maximally spinning Kerr PBHs with $a_* = 0.999$ to assess the extent to which the bounds can be strengthened, as we expect only mild improvements for moderately rotating PBHs. The results are shown in Fig.~\ref{fig:spin_fpbhlim}, where we find that the bounds improve by a factor of $\sim7$ relative to those in Fig.~\ref{fig:res}. These results indicate that the constraints presented in this paper are conservative. We also note that for the case of transit of Kerr PBHs (which we do not consider in this work), the emission spectrum depends on the observer angle with respect to the spin axis~\cite{Perez-Gonzalez:2023uoi}. This requires a more detailed analysis based on the geometry.
\section{Comparison to other (low-energy) neutrino limits}
\label{appsec:nu_lim}
\begin{figure}
\centering
\includegraphics[width=0.5\textwidth]{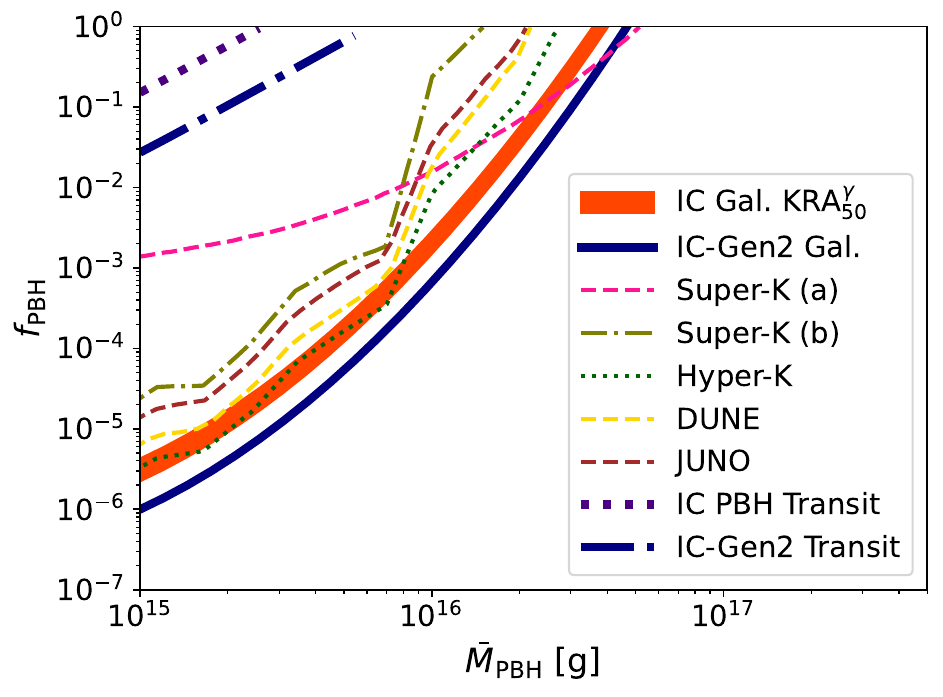}
\caption{\label{fig:nu_comp}Comparison of limits derived in this work with previous low-energy neutrino limits for from Super-K (a)~\cite{Dasgupta:2019cae}, Super-K (b), Hyper-K, DUNE and JUNO~\cite{Bernal:2022swt}. The limits are derived for Schwarzschild PBHs with a log-normal extended mass function with $\sigma = 1.0$ in accordance with the previous work. We show our most stringent limit using current generation telescope IceCube for the KRA$_{50}^\gamma$ template and the projection of the same for IceCube-Gen2.
}
\end{figure}
In this work, we highlight for the first time the potential of high-energy neutrinos to place bounds on $f_{\rm PBH}$. Existing constraints on $f_{\rm PBH}$ from low-energy neutrino observations at Super-Kamiokande have been presented in Refs.~\cite{Dasgupta:2019cae} and~\cite{Bernal:2022swt}. In this section, we compare our high-energy neutrino bounds with those derived from low-energy observations. Since Refs.~\cite{Dasgupta:2019cae} and~\cite{Bernal:2022swt} report their results for the relevant mass range only for a log-normal mass function, we adopt the same distribution here instead of the GCC mass function used throughout this work. For consistency, we fix $\sigma = 1$ and $a_* = 0$.

Our results are shown in Fig.~\ref{fig:nu_comp}. We find that bounds derived from the diffuse high-energy neutrino flux, using current and upcoming detectors, are competitive with both existing and projected low-energy neutrino constraints at larger PBH peak masses $\bar{M}_{\rm PBH}$. While low-energy neutrino observatories are, in principle, sensitive to more massive (and thus colder) PBHs, probing mass ranges less suppressed by the mass function, high-energy neutrino telescopes benefit from significantly larger (roughly $3$ orders of magnitude larger) effective areas, which partially compensates for this effect. At lower $\bar{M}_{\rm PBH}$, our bounds become more stringent, as the low mass tail of the PBH distribution is enhanced, leading to a stronger high-energy neutrino signal.
\end{document}